# Broadband Fourier-Transform Optical Photothermal Infrared Spectroscopy and Imaging

Aleksandr Razumtcev[1,2,*], Gwendylan A. Turner[1,2,3], Sergey Zayats[4], Ferenc Borondics[5], Aris Polyzos[3], Garth J. Simpson[2,+], and Hans A. Bechtel[1,+]

[1]Advanced Light Source Division, Lawrence Berkeley National Laboratory, Berkeley, CA 94720

[2]Department of Chemistry, Purdue University, West Lafayette, IN 47907

[3]Molecular Biophysics and Integrated Bioimaging Division, Lawrence Berkeley National Laboratory, Berkeley, CA 94720

[4]Photothermal Spectroscopy Corp., Santa Barbara, CA 93101

[5]Synchrotron SOLEIL, Saint Aubin, France 91190

[+]Co-corresponding authors

[*]Current affiliation: Molecular Foundry, Lawrence Berkeley National Laboratory, Berkeley, CA 94720

**Abstract:** Optical photothermal microscopy is a powerful, emerging method that overcomes the diffraction limit in infrared hyperspectral imaging by utilizing a visible probe laser beam to detect local temperature-induced modulation at the visible diffraction limit. However, the spectral range of this technique has been limited by the tuning range of infrared sources, which is generally restricted to the fingerprint window with commercially available quantum cascade lasers. In this work, ultra-broadband synchrotron infrared radiation was used for infrared photothermal imaging and spectroscopy, spanning the entire mid-infrared range. Both optical- and fluorescence-detected photothermal modalities were performed, demonstrating improved spectral range when compared to optical photothermal microscopy using commercial sources and improved spatial resolution when compared to synchrotron micro-spectroscopy measurements. Following these initial validation studies, synchrotron Fourier-transform optical photothermal infrared spectroscopy (FT-OPTIR) in combination with synchrotron micro-spectroscopy measurements were used to differentiate cells in mouse brain tissue sections.

**Introduction**

Infrared (IR) micro-spectroscopy has become one of the most widely used methods for chemical identification and characterization of small samples due to its high specificity and non-invasive nature. IR spectroscopy in the so-called "fingerprint region" of the electromagnetic spectrum often allows unambiguous discrimination between chemical compounds in a variety of samples, including living cells and organisms, with every major class of biochemical building blocks having distinct spectral signatures (e.g., amide absorption bands for the proteins).[1] However, diffraction and the long wavelengths of IR light have traditionally limited the spatial resolution of infrared micro-spectroscopy to the µm scale, hindering measurements of sub-cellular components and other nanoscale heterogeneity.

Recently, the combination of IR spectroscopy with scanning probe-based techniques has enabled near-field IR spectroscopy well beyond the diffraction limit and into the nanoscale. Both scattering type, scanning near-field optical microscopy (s-SNOM)[2–4] and thermal-expansion based photothermal atomic force microscopy (AFM-IR)[5,6] modalities provide tip-limited spatial resolution less than 20 nm and have been applied to a wide variety of samples. However, these scanning probe techniques confine hyperspectral analysis to AFM-compatible samples, thereby limiting potential applications, particularly toward hydrated biological materials. Furthermore, AFM-based nanospectroscopy measurements are time-consuming, a challenge that is exacerbated for the large fields-of-view (FoV) routinely necessary for biological imaging. These factors have resulted in a measurement gap in which there remains a need for a rapid, non-tip-based method to chemically characterize samples with a spatial resolution superior to IR microspectroscopy.

Optical photothermal mid-IR (O-PTIR) microscopy has gained prominence over the last decade as a far-field optical method that overcomes the diffraction limit of mid-IR light.[7–10] O-PTIR has been applied for chemically-specific imaging in a variety of fields, including tissue analysis, live-cell imaging, pharmaceutical materials characterization, and microplastics detection among others.[11–15] In O-PTIR, the spatial resolution is dictated by the optical resolution of a visible probe laser, which is typically around 500 nm for commercial systems. The most common mechanism for transducing the photothermal signal is based on temperature-induced variations of the local refractive index. A recent report demonstrated that the spatial resolution in O-PTIR imaging can go beyond the diffraction limit of visible light by probing temporal dynamics of photothermal relaxation, reaching a spatial resolution below 200 nm in live cell imaging.[16] Separately reported advances in O-PTIR instrumentation have also demonstrated high-speed analysis at video rate by wide-field imaging or using galvo mirrors for fast beam scanning.[15,17] Despite these advances, achievable sensitivities have been limited because the refractive index of a material is a relatively weak function of temperature.[18]

Fluorescence-based far-field detection of IR photothermal imaging (F-PTIR) has recently been shown to considerably enhance the sensitivity of O-PTIR by two orders of magnitude by leveraging the temperature-sensitivity of fluorescence emission.[19,20] Unlike other techniques that use intramolecular transition double-resonance strategies to exploit the brightness and sensitivity of fluorescence emission, [21,22,23] F-PTIR relies on the detection of accompanying temperature changes within the bath-matrix adjacent to the fluorophore. Thus, F-PTIR is sensitive to a variety of molecules near the fluorescent tag and not only the fluorescent tag itself. By further leveraging the specificity of fluorescence labeling, F-PTIR can be used to extract chemical information on specific cell types and sub-cellular organelles.[20]

Despite having superior spatial resolution when compared to diffraction-limited FTIR imaging, the spectral range of optical-based PTIR techniques remain limited due to the availability of commercial IR radiation sources. Because of the low temperature sensitivity of photothermal signal transducers, such as refractive index, a local transient temperature change of several ºC is usually required to produce a detectable photothermal response.[18] Rapid thermal dissipation of temperature gradients drives the use of high-power short pulses of mid-IR light, such as those produced by QCLs. However, the range of a single QCL chip usually covers less than 500 cm$^{-1}$ in the mid-IR region, and not more than 4 chips are installed into commercially available modules. Furthermore, currently available commercial QCL sources have a limited spectral range, making it difficult to access frequencies above 3000 cm$^{-1}$ and below 800 cm$^{-1}$,[24] although a few research prototypes have overcome this limitation in the THz region.[25–28]

Synchrotron-generated IR radiation, on the other hand, offers a unique combination of high spectral irradiance and ultrabroad bandwidth.[29] With a spectral brilliance nearly three orders of magnitude higher than thermal IR sources used in commercial FTIR spectrometers, and a spectral bandwidth spanning the far-IR, mid-IR, near-IR and beyond, synchrotron IR radiation is routinely used for IR microspectroscopy[30–33] and IR nanospectroscopy measurements.[34–39] In this work, we overcome the bandwidth limitations of laboratory-based infrared sources by using broadband synchrotron IR radiation for optical photothermal spectroscopy and imaging. Broadband photothermal spectroscopy is achieved by employing a Michelson interferometer approach followed by Fourier transformation (FT) of the collected interferogram. Both synchrotron O-PTIR and F-PTIR are successfully employed to enable broadband high-resolution IR imaging of model systems and fluorescently labeled fixed mouse brain tissues.

**Results**

Two complementary beampaths were constructed for broadband synchrotron-based PTIR at the Advanced Light Source (ALS) beamlines 1.4 and 2.4 to assess the performance of two different signal generation strategies: fluorescence-based F-PTIR and refractive index-based O-PTIR, respectively. Both instruments are described in detail in the Methods section. In brief, the synchrotron O-PTIR system (**Fig. 1a**) was built around a commercial O-PTIR microscope (mIRage-LS, Photothermal Spectroscopy Corp.) with a flip mirror installed in the beampath to switch between synchrotron and QCL modalities, whereas the F-PTIR system was built around a modified commercial infrared microscope (Nicolet Nic-plan) equipped with a photomultiplier tube for detection of fluorescence emission in an epi-configuration (**Fig. 1b**).

Prior to entering either microscope, the synchrotron beam was directed to a custom-built step-scan interferometer and an optical chopper. Although, ALS synchrotron radiation is inherently pulsed (60 ps), it also has a high repetition rate (500 MHz). This high repetition rate is too fast for sufficient thermal relaxation between pulses, and thus modulating the light at a slower repetition rate is necessary in order to observe a thermal gradient. Lock-in detection of the O-PTIR or F-PTIR signal was performed at each position of the interferometer moving mirror in a step-scan approach. Fourier-transformation of the PTIR signals as a function of interferometer mirror position produced localized IR absorption spectra over a range of IR frequencies between 500 and 4000 $cm^{-1}$, as determined by the combination of transmission ranges of the Ge window and ZnSe beamsplitter.

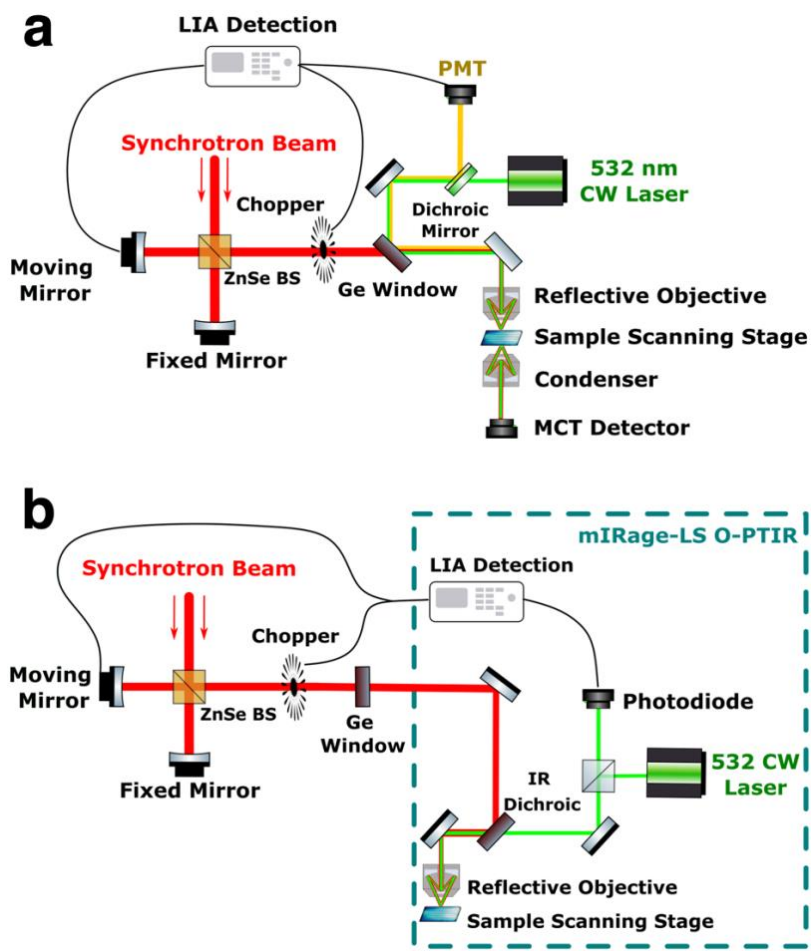

**Figure 1.** Instrument diagrams for PTIR systems integrated into the ALS beamlines. (a) The instrument used for O-PTIR spectroscopy was built around a commercial O-PTIR system. Broadband synchrotron infrared light passed through an interferometer before entering the microscope. Back-scattered 532 nm probe light was demodulated at the chopper frequency at each position of the interferometer to extract the amplitude of the photothermal modulation. (b) – The instrument used for F-PTIR spectroscopy and imaging had a similar design, but epi-detected fluorescence emission was used as the photothermal signal reporter. This instrument supported simultaneous transmission FTIR imaging.

Synchrotron O-PTIR interferograms and resulting broadband IR absorption spectra are shown in **Figure 2** for two model polymer materials (polystyrene and polyethylene terephthalate, or PET) overlaid with their respective ATR FTIR and QCL O-PTIR spectra. Fourier-transform broadband O-PTIR (FT-OPTIR) recovered all major fingerprint region absorption features of the studied polymers with the nominal spectral resolution of 8 cm$^{-1}$ corresponding to 0.0625 cm total distance traveled by the moving mirror. Notably, low-frequency bending vibrations of the aromatic rings for both polymers (536 cm$^{-1}$ for polystyrene and 728 cm$^{-1}$ for PET) were clearly resolved.

The signal-to-noise ratio (SNR) for the most prominent peaks was 43 for the polystyrene peak at 1278 cm$^{-1}$ and 52 for the PET peak at 682 cm$^{-1}$. The SNR was calculated for a mean spectrum that was averaged over 10 individual spectra with 1 second integration time per interferometer position. Several factors likely limited the observed SNR in FT-OPTIR measurements. Physical limitations on the optical chopper used in these experiments capped IR modulation frequencies to < ~1 kHz, leading to increased 1/f measurement noise. Optical losses, mechanical vibrations, and non-ideal beam overlap in the custom interferometer may have also contributed to spectral degradation, particularly in the higher energy portion of the spectrum. Finally, the relatively low IR power of the synchrotron source itself (~4 mW integrated across the whole spectral range) impacted the SNR. Most of these factors have the potential for improvement in future implementations (e.g., by using more advanced interferometers, correcting IR beam shape with adaptive optics, and planned upgrades to synchrotron infrared beamlines).

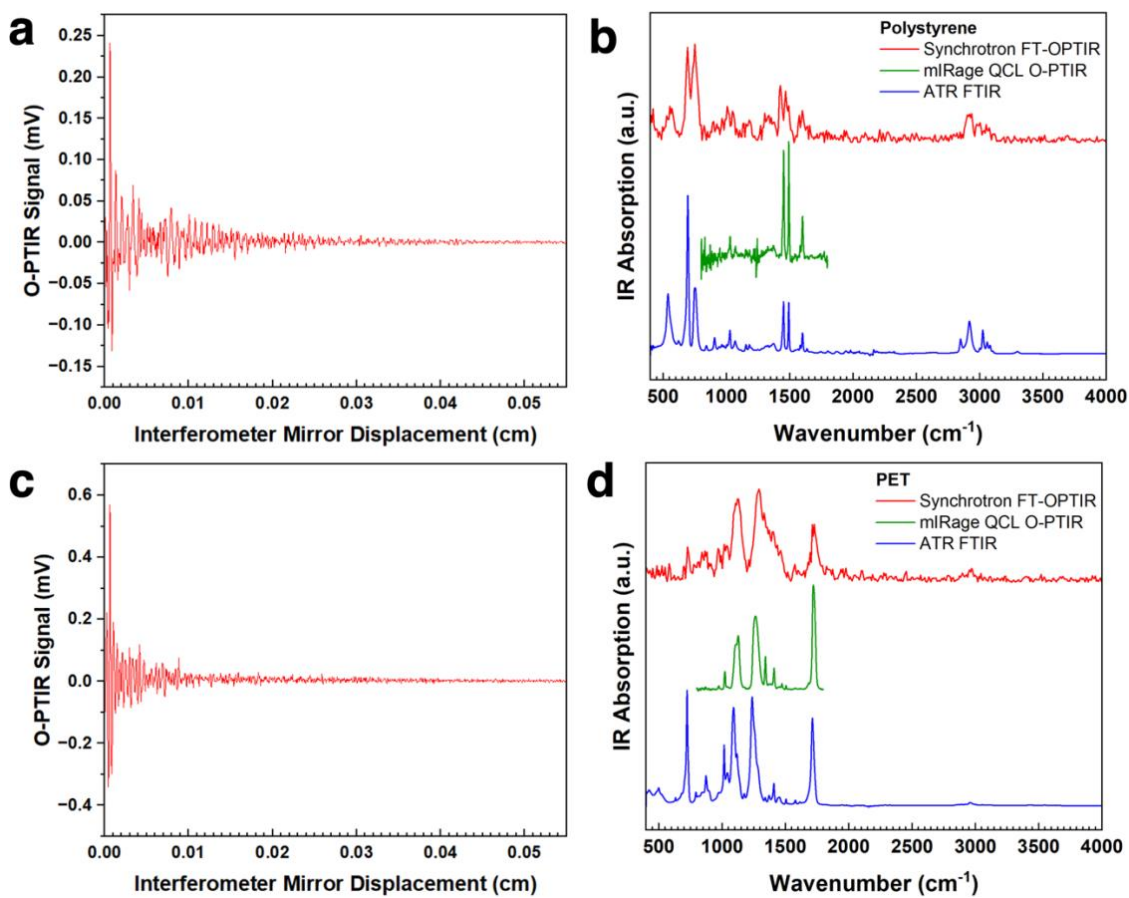

**Figure 2.** Synchrotron FT-OPTIR spectroscopy of model polymer materials: polystyrene (a and b) and PET (c and d). An interference pattern in the O-PTIR signal was observed as a function of the interferometer mirror position. The shown interferograms were produced by bandpass filtering and apodization of the raw FT-OPTIR signal. Synchrotron FT-OPTIR significantly extends the range of optical photothermal spectroscopy in both high- and low-frequency parts of the mid-IR spectrum.

Following the proof-of-concept synchrotron FT-OPTIR spectroscopy measurements, we evaluated the improvement in spatial resolution of synchrotron F-PTIR over conventional synchrotron FTIR microscopy (**Figure 3**). An optical microscope utilizing fluorescence-based F-PTIR detection modality was integrated into an existing ALS FTIR microspectroscopy beamline. The instrument supported simultaneous broadband FT-FPTIR and FTIR imaging by detecting epi-fluorescence emission for the former and transmitted IR intensity for the latter. Rhodamine-6G associated silica gel particles were used for initial measurements due to their high IR absorption and easily interpretable mid-IR spectrum. Single-point synchrotron FT-FPTIR spectroscopy maintained an identical spectral range (**Fig. 3b**) with enough spectral fidelity to clearly resolve the absorption peak of silica gel at 1150 cm$^{-1}$. It should be noted that, consistent with previously published results,[19,20] no vibrational features were detected that were associated with the fluorophore.

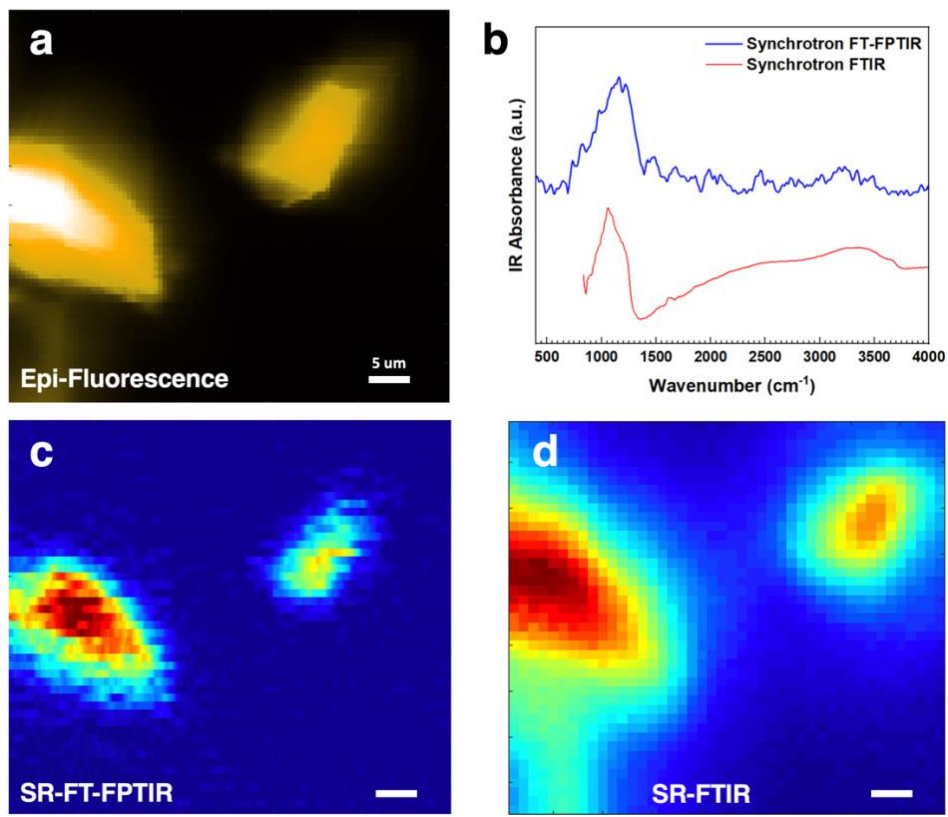

**Figure 3.** A performance comparison between synchrotron (SR) FT-FPTIR (c) and FTIR (d) microspectroscopy demonstrates a spatial resolution improvement achieved by the photothermal modality. Panel (a) is an epi-fluorescence image of the studied FoV for reference. Panel (b) shows representative single-pixel mid-IR absorption spectra for both imaging modes.

A clear spatial resolution improvement can be seen in the FT-FPTIR image as compared to its FTIR counterpart. The spatial resolution for FT-FPTIR and FTIR was calculated using a knife edge method for an FoV containing a single small particle (data shown in the **Supporting Information**). A spatial resolution of 1.5 µm was achieved with synchrotron FT-FPTIR, as compared to 8.1 µm for the diffraction-limited FTIR imaging of the same FoV, resulting in a 5.4-fold resolution improvement at 1150 cm$^{-1}$. We note that the silica gel particles were not perfect step edges due to their geometry and thickness, and thus, the spatial resolution of the system for FT-FPTIR is likely higher than stated. Furthermore, the spatial resolution determination in synchrotron FT-FPTIR measurements was pixel count limited with a single pixel size of 1 µm to decrease the acquisition time. A significant resolution improvement is expected to be achieved in instruments with counter-propagating visible and infrared beams using high numerical aperture objectives for both incident beams.

Following the imaging experiments on model fluorescent particles described above, broadband synchrotron FT-FPTIR imaging was performed on fixed thin mouse brain tissue sections labeled by immunofluorescence. The measurements were performed in the striatum region of the brain that was previously connected to neuronal death in Huntington's disease patients.[40] Cells were stained by the NucSpot 555® fluorescent dye, which selectively targets cell nuclei. The results of broadband photothermal imaging are shown in **Figure 4** alongside representative epi-fluorescence, bright field, and diffraction-limited hyperspectral FTIR microscopy results of the same FoV. As can be seen in **Fig. 4**, broadband FT-FPTIR enabled selective high-resolution mapping of IR absorption of cell nuclei within the tissue, while the resolution and contrast between cell bodies and surrounding tissue were mostly lost in the conventional FTIR measurements. The FTIR absorption map at 1655 cm$^{-1}$ in **Fig. 4** corresponds to one of the major fingerprint region signatures of proteins. FTIR maps at other major mid-IR absorption peaks of biomolecules (including the protein Amide II peak, the nucleic acids phosphate absorption peak, and lipid peaks) are shown in the **Supporting Information**. Because of the ubiquitous distribution of major biomolecule classes within biological systems, these spectral bands produced little contrast that would enable the identification of cell bodies.

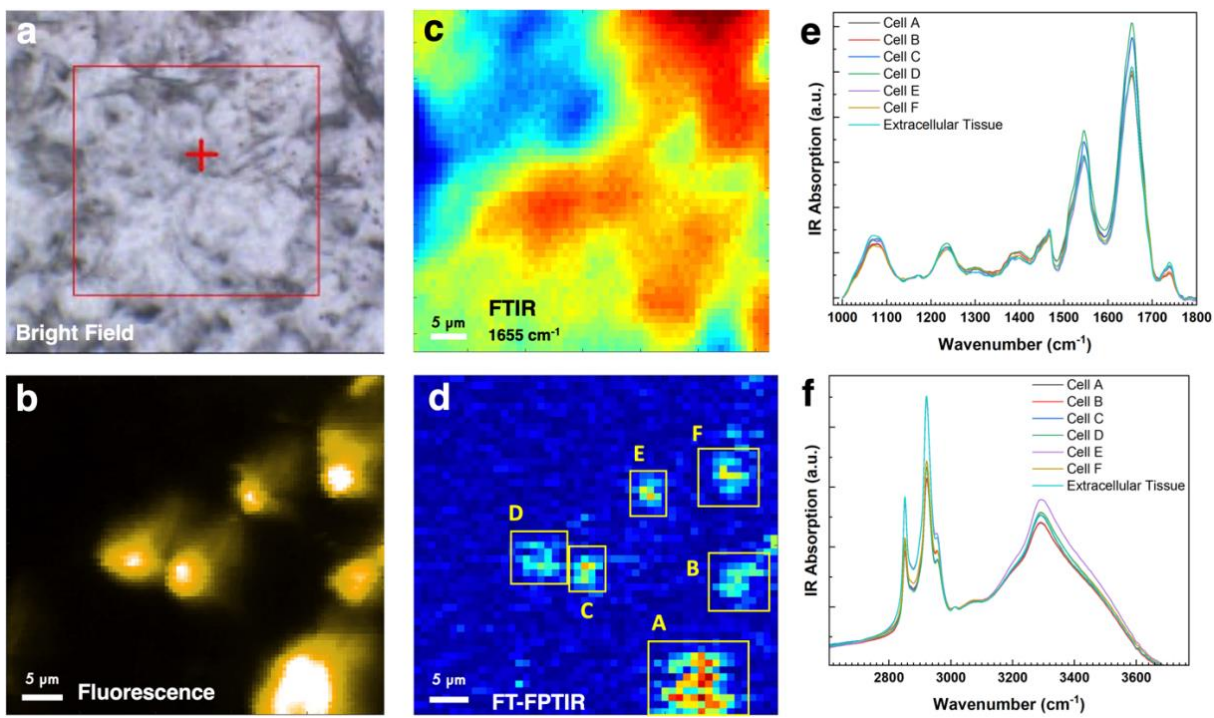

**Figure 4.** Cell-specific broadband FT-FPTIR chemical imaging of a fluorescently-labeled mouse tissue section. The integrated IR absorption FT-FPTIR image shown in (d) only has features coinciding with fluorescence-labeled cell nuclei that can be seen in the fluorescence image shown in (b). At the same time, the contrast between cell bodies and the surrounding tissue is mostly lost in the FTIR results (c), or in the bright-field image (a). FT-FPTIR guided cell-specific synchrotron FTIR spectroscopy results for the extracellular tissue, cell bodies, and a single cell (yellow box in d) for different parts of the mid-IR region are shown in (e) and (f)

Due to the higher spatial localization of IR absorption signal, broadband FT-FPTIR absorption maps were used as a guide to create masks for hyperspectral synchrotron FTIR dataset and extract full range IR absorption information for single cell bodies. The results shown in **Fig. 4 e-f** show the variations in IR absorption spectra between the extracellular tissue and the signal arising from individual cells labeled on the FT-FPTIR absorption map in **Fig. 4d**. The variations were relatively minor in the fingerprint region (1000 – 1800 cm$^{-1}$). However, a larger relative ratio of the lipid absorption peak at 1750 cm$^{-1}$ to the protein amide I spectral signature in the extracellular region (0.16 vs. an average of 0.12 across the cell bodies, see **Supporting Information**) might indicate a lower relative concentration of lipids within the cell bodies, which is consistent with the striatum being a lipid-rich region of the brain. This difference was significantly more pronounced in the higher-frequency region. The ratio of the lipid CH-stretch peak intensities (2850 and 2920 cm$^{-1}$) to the intensity of the protein amide A peak (3300 cm$^{-1}$) (**Supporting Information**) was significantly higher for the surrounding tissue (1.76) as compared to the signal originating from any of the individual cells (1.36). The low lipid concentration within brain cell nuclei can be additionally visualized by plotting a ratio map of the lipids to proteins CH-stretch region peaks that is also provided in the **Supporting Information**.

## Discussion

The FT-OPTIR and FT-FPTIR results presented here demonstrate the potential of broadband far-field photothermal spectroscopy. The collected synchrotron FT-OPTIR spectra of model polymer and inorganic materials continuously cover the spectral range between 500 and 4000 cm$^{-1}$, extending the range substantially over that currently accessible with commercial QCL arrays in both higher and lower frequency regimes. However, it is important to note that the low-frequency cut-off was not dictated by the synchrotron source bandwidth, but rather by the optical transparency of the optical components (ZnSe beamsplitter and Ge window) and could be extended into the far-IR by using materials with higher transparency.

The potential to perform photothermal IR imaging in the far-IR region of the spectrum is particularly appealing due to the lack of commercially available high-flux laser sources covering this region. For a model polystyrene sample studied in this work, the lowest frequency absorption peak resolved when using a commercial QCL was the vibrational mode at 1452 cm$^{-1}$, while the collected broadband FT-OPTIR spectrum contained resolvable transitions at 755 cm$^{-1}$, 686 cm$^{-1}$, and 541 cm$^{-1}$. Furthermore, the signal detection in FT-OPTIR spectroscopy relied exclusively on the back-scattered visible light, supporting IR microscopy of thick samples with negligible IR transmissivity. Photothermal detection also supports potential improvements in sensitivity in the far-IR region over routine FTIR measurements because commonly used mercury-cadmium-telluride (MCT) detectors experience dramatic reduction in sensitivity in this region.

In addition to the considerable extension of the accessible spectral range discussed above, IR photothermal far-field spectroscopy arguably holds promise for single-cell chemically-specific hyperspectral imaging, enabled by the resolution improvement over diffraction-limited mid-IR imaging. Side-by-side comparisons in this work confirm a 5.4-fold resolution improvement for FT-FPTIR with fluorescence detection over FTIR transmission imaging in the same field-of-view in the fingerprint region around 9 µm wavelengths. Furthermore, because of the shorter Raleigh length for visible light, F-PTIR imaging has an improved resolution in the z-direction along the optical axis for optical sectioning, as evidenced by the lack of FT-FPTIR signal from out-of-focus features of the large silica gel particle that are observed in the fluorescence and FTIR images (**Fig. 3**).

One notable advantage of O-PTIR techniques directly lies on their reliance on visible light for detection, facilitating analysis of thick or IR-absorbing samples. The limited penetration depth of IR radiation through some materials typically restricts transmission IR measurements, such as water, which at thicknesses greater than 10 µm readily absorbs incident IR beams, and generally renders measurements in the water absorption region impractical. In contrast, the visible probe beam employed in optical PTIR techniques can traverse millimeter-thick water layers with negligible attenuation, removing this constraint. Moreover, instrumentation that is configured for counter-propagation can illuminate samples with IR light from the substrate side, enabling efficient delivery of mid-IR light to surface-adherent biological specimens.

In addition to enhancing the measurement sensitivity, F-PTIR also offers targeted IR imaging in complex biological samples, leveraging a vast library of highly specific fluorescent dyes. In this work, brain tissue sections were imaged from mice with Huntington's disease (HD). HD is a currently untreatable lethal neurodegenerative disease.[41] One of the known HD mechanisms leading to the loss of cognitive function of the brain is neuronal death in the striatum region.[40] Investigating the structural and metabolical changes in neurons leading to their death clearly benefits from targeted characterization of this particular cell type in brain tissue, which can be challenging to achieve with commercial FTIR microscopes due to the lack of specificity and insufficient spatial resolution.[42,43] Herein, as a proof-of-concept demonstration of targeted high-resolution IR imaging of cells in tissue sections, FT-FPTIR analysis was performed on fixed sections labeled with a cell nucleus-specific stain. FT-FPTIR signal was only generated in the regions labeled by fluorescence and therefore was only observed in pixels coinciding with the locations of cell nuclei in the epi-fluorescence image of the same FoV (**Fig. 4**). No contrast between the cell bodies or nuclei and the surrounding tissues was detected at any of the major absorption peaks in FTIR images due to the uniform distribution of major classes of protein and lipid molecules throughout the whole tissue. Using the cell-specific FT-FPTIR result to guide the FTIR analysis enabled extraction of the FTIR signal corresponding to a single cell body, as well means to spatially differentiate the signal originating from the cell nuclei from the surrounding tissue.

Despite encouraging results obtained in this proof-of-concept demonstration, the FT-PTIR implementation presented herein has room for improvement in future implementations. Relatively low signal-to-noise in the collected interferograms and recovered spectra required prolonged averaging at each interferometer position, resulting in long single-spectrum acquisition times (up to 10 minutes). Such a long single-pixel acquisition time prevented full FoV hyperspectral FT-FPTIR imaging in the current implementation. The contrast in high-resolution FT-FPTIR images was due to the photothermal effect caused by the absorption of the broadband beam as a whole, while the spectral information was recovered by conventional low-resolution FTIR imaging of the same FoV. As discussed in the results section, the SNR in FT-PTIR measurements is expected to be significantly improved in future work by using improved commercial interferometers and synchrotron beam shaping. Additionally, high-speed modulation with fast optical choppers should, in principle, result in significant suppression of the *1/f* measurement noise.

In summary, interferometry-based broadband synchrotron FT-PTIR spectroscopy and imaging are demonstrated. FT-PTIR is shown to provide a substantial extension of the accessible IR spectral range in optical photothermal measurements and enable sub-micron chemical imaging in the fingerprint region. Furthermore, fluorescence-based detection opens additional pathways to perform targeted vibrational spectroscopic imaging in complex biological samples, as demonstrated on mouse brain tissues labeled with a nucleus-specific fluorescence dye. The methodology discussed in this work has the potential to revolutionize synchrotron infrared science by extending far-field synchrotron-based infrared microspectroscopy beyond the diffraction limit. Additionally, these proof-of-concept results suggest that optical photothermal imaging can be realized with broadband sources such as supercontinuum and ultra-broadband IR pulsed lasers. Further technological improvement of such light sources has the potential to open the pathways to developing high-speed sensitive chemically-specific imaging modalities not requiring access to synchrotron facilities.

**Materials and methods**

*ALS synchrotron beam parameters*

All measurements were performed at the Advanced Light Source (ALS) Beamlines 1.4 and 2.4 (Lawrence Berkeley National Laboratory, Berkeley, CA). During normal user operations, the ALS operates at 1.9 GeV with 500 mA current in top-off mode, producing light spanning from the far-IR to hard X-ray. At the IR beamlines, the X-ray and UV radiation is removed prior to being delivered to the endstations with a series of aluminum and gold-coated mirrors. A diamond window separates the ultra-high vacuum of the storage ring from the beamline's ambient conditions. Approximately 0.5 mW of IR radiation, integrated between 500-5000 $cm^{-1}$, was used for the O-PTIR measurements at Beamline 2.4, and approximately 1 mW of IR radiation was used for the F-PTIR measurements at Beamline 1.4.

*Synchrotron PTIR spectroscopy and microscopy instrumentation*

The instrument for broadband FT-OPTIR spectroscopy was a commercial O-PTIR microscope (mIRage-LS manufactured by Photothermal Spectroscopy Corp, Santa Barbara, CA, USA), which was modified in collaboration with the vendor to facilitate beamline integration. Specifically, a flip mirror was installed into the beampath for convenient switching between the synchrotron and QCL operation modes. Prior to entering the instrument, the IR beam was passed through a custom-built step-scan Michelson interferometer using a ZnSe beamsplitter, where the moving mirror was placed on a linear nanopositioning stage (Aerotech ANT95-L). Except for the interferometer components and a Ge window placed in the beampath to reject the visible component of the synchrotron beam, no additional changes were required in the optics in the mIRage-LS system. A 532nm CW laser was used as a probe beam, and a Si photodiode was used to detect the scattered light and extract the O-PTIR signal. The mIRage-LS was switched to its normal QCL-based operation mode to collect reference QCL O-PTIR spectra on the same spot as the SR based measurements were done.

The microscope used for synchrotron F-PTIR measurements was built around a Nicolet Nic-plan IR microscope. The side port of the microscope was used to couple in a fluorescence excitation light coming from a 532nm laser diode (CPS 532, Thorlabs). The fluorescence emission was detected in an epi-configuration and filtered from the excitation light by a 550 shortpass dichroic mirror (Thorlabs), and a combination of fluorescence filters (532 notch filter from Edmund Optics, and 550 long-pass filter from Thorlabs). A liquid nitrogen-cooled MCT/A detector was used to detect transmitted IR beam for diffraction-limited FTIR imaging. The synchrotron IR beam and the fluorescence excitation beam were combined on a germanium window and focused on the sample with a 32x 0.65 NA reflective objective (SpectraTech Inc). The average visible excitation laser power was ~0.5 mW. The sample was raster-scanned using a Prior Scientific Instruments XY microscope stage H101A to generate 50x50 pixel images with 50x50 µm fields of view with a single pixel dwell time of 1 ms for fluorescence imaging and 500 ms for FT-FPTIR imaging.

In both cases, the synchrotron IR beam was chopped by a mechanical chopper (Stanford Research Systems), placed before the interferometer, at frequencies ranging from 200 to 700 Hz. The PTIR signal at each interferometer mirror position was extracted by using a lock-in amplifier (Zurich Instruments MFLI for the O-PTIR system and Stanford Research Systems SR810 for the F-PTIR system). The signal integration time per interferometer position varied between 200ms and 1s. The interferometer moving mirror traversed a total distance of 0.0625 $cm^{-1}$ with 1000 intermediate steps.

*Data collection and analysis*

For the O-PTIR system, the signal was demodulated at the chopper modulation frequency using a Zurich Instruments lock-in amplifier at each position of the interferometer to produce experimental O-PTIR interferograms. The interferograms were high-pass filtered to remove low-frequency baseline drift using the OriginLab data analysis software package and then apodized with the Happ-Genzel function. The processed interferograms were then Fourier-transformed and the amplitude of the transform result was extracted and plotted as an IR absorption intensity.

In the F-PTIR system, the raw PMT (Hamamatsu) signal was pre-amplified (Stanford Research SR560) and demodulated at the chopper frequency using an SRS SR810 lock-in amplifier. The signal was then digitized using an AlazarTech ATS9462 digital oscilloscope card. A custom MATLAB algorithm was used to control the microscope stage and reconstruct 50x50 pixel images for a 50x50 µm fields-of-view.

*FTIR spectroscopy and hyperspectral imaging*

Reference FTIR spectra for all materials were collected using a Nicolet iS50 FTIR spectrometer (Thermo Fisher Scientific). The spectra for polystyrene and PET were acquired in the attenuated total reflectance (ATR) mode using the built-in thermal IR source in the range between 400 and 5000 cm$^{-1}$ with a 4 cm$^{-1}$ spectral resolution.

Synchrotron FTIR microspectroscopy of the silica gel particles and brain tissue sections was conducted at the ALS Beamline 1.4 using a Nicolet iS50 spectrometer (Thermo Fisher Scientific) and a Nicolet Nic-plan microscope. The IR absorption maps were collected with a 1 µm step for both the x and y axes, and each spectrum was collected in the 600 – 5000 cm$^{-1}$ spectral range with 1 cm$^{-1}$ spectral resolution.

*Sample preparation*

Thin films of PET and polystyrene were provided by Photothermal Spectroscopy Inc. R6G-associated silica gel particles were prepared by mixing 100 mg of silica gel (60-200 µm particles, SiliCycle) and 5 mg of R6G in 10 mL of deionized water (1.05 mM R6G concentration) and air drying the extracted particles. Silica gel was ground with a mortar and pestle prior to labeling to reduce the average particle size.

Coronal mouse brain sections were cryo-embedded in Optical Cutting Temperature Compound (OCT) at -80° C and sliced into 15 µm sections using a Leica Cryostat set at -14° C. The resulting sections were fixed onto CaF$_2$ slides (Crystan, UK. Part #CAFP25-1) with 4% paraformaldehyde for 20 minutes at room temperature, washed three times in phosphate-buffered saline solution (pH 7.2), and permeabilized. Lipofuscin autofluorescence was quenched using Trueblack (Biotium, USA. Part#23007) in 70% ethanol, and NucSpot® 555 (Biotum) was applied to the tissue at a 30,000 fold dilution for nuclei staining. Tissues were rinsed with phosphate buffered saline, re-fixed with 4% PFA, and kept frozen at -4˚C until F-PTIR measurements.

**Acknowledgements**

The authors gratefully acknowledge funding for the present work from the National Science Foundation (CHE-2004046, CHE-2305178, CHE-2320751, CIF-1763896) and the NSF Center for Bioanalytic Metrology (IIP-1916691).

This research used resources of the Advanced Light Source, which is a DOE Office of Science User Facility under contract no. DE-AC02-05CH11231. Aleksandr Razumtcev was supported in part by an ALS Doctoral Fellowship in Residence.

The authors are grateful to Photothermal Spectroscopy Corp. for providing the mIRage-LS instrument for the experiments and their engineering team for assistance with the modification and installation of the instrument at the ALS.

**Author contributions**

All authors contributed to the design, data acquisition, analysis, and writing of this work.

**Competing interests**

The authors declare the following competing interests:

Sergey Zayac is an employee of Phothermal Spectroscopy Corp.

The remaining authors declare no competing interests.